%% file: qedmod_v5.tex
\documentclass[twocolumn,aps,pra,floatfix,superscriptaddress,notitlepage]{revtex4-1}
\usepackage[utf8]{inputenc}
\usepackage[OT1]{fontenc}
\usepackage{amsmath}
\usepackage{amsfonts}
\usepackage{amssymb}
\usepackage{bm}
\usepackage{graphicx}
\usepackage[left=2cm,right=2cm,top=2cm,bottom=2cm]{geometry}
\usepackage{longtable,booktabs}
\allowdisplaybreaks

\usepackage{dcolumn}
\usepackage{siunitx}
\usepackage{multirow}

\usepackage{color}
\definecolor{BLUE}{rgb}{0.0,0.0,1.0}

\usepackage{soul}
\soulregister\cite7
\soulregister\ref7

\newcommand{\veps}{\varepsilon}
\newcommand{\balpha}{\bm{\alpha}}

\newcommand{\br}{\bm{r}}

\newcommand{\bp}{\bm{p}}

\newcommand{\be}{\begin{eqnarray}}
\newcommand{\ee}{\end{eqnarray}}

\begin{document}

\title{Model-QED operator for superheavy elements}

\author{A.~V.~Malyshev}
\affiliation{Department of Physics, St.~Petersburg State University, Universitetskaya 7/9, 199034 St.~Petersburg, Russia  
\looseness=-1}

\author{D.~A.~Glazov}
\affiliation{Department of Physics, St.~Petersburg State University, Universitetskaya 7/9, 199034 St.~Petersburg, Russia  
\looseness=-1}

\author{V.~M.~Shabaev}
\affiliation{Department of Physics, St.~Petersburg State University, Universitetskaya 7/9, 199034 St.~Petersburg, Russia  
\looseness=-1}

\author{I.~I.~Tupitsyn}
\affiliation{Department of Physics, St.~Petersburg State University, Universitetskaya 7/9, 199034 St.~Petersburg, Russia  
\looseness=-1}

\author{V.~A.~Yerokhin}
\affiliation{Peter the Great St.~Petersburg Polytechnic University, Polytekhnicheskaya 29, 195251 St.~Petersburg, Russia  
\looseness=-1}

\author{V.~A.~Zaytsev}
\affiliation{Department of Physics, St.~Petersburg State University, Universitetskaya 7/9, 199034 St.~Petersburg, Russia  
\looseness=-1}


\begin{abstract}

The model-QED-operator approach [Phys. Rev. A {\bf 88}, 012513 (2013)] to calculations of the radiative corrections to binding and transition energies in atomic systems is extended to the range of nuclear charges $110 \leqslant Z \leqslant 170$. The self-energy part of the model operator is represented by a nonlocal potential based on diagonal and off-diagonal matrix elements of the \textit{ab initio} self-energy operator with the Dirac--Coulomb wave functions. The vacuum-polarization part consists of the Uehling contribution which is readily computed for an arbitrary nuclear-charge distribution and the Wichmann--Kroll contribution represented in terms of matrix elements similarly to the self-energy part. Performance of the method is studied by comparing the model-QED-operator predictions with the results of \textit{ab initio} calculations. The model-QED operator can be conveniently incorporated in any numerical approach based on the Dirac--Coulomb--Breit Hamiltonian to account for the QED effects in a wide variety of superheavy elements.

\end{abstract}


\maketitle


\section{Introduction \label{sec:0}}

Recent advances in the synthesis and experimental investigation of electronic-structure and chemical properties of superheavy elements as well as the prospects for further progress in this field~\cite{Oganessian:2016:901, Oganessian:2017:023003, Nazarewicz:2018:537, Giuliani:2019:011001, Dullmann:2019:587, Oganessian:2019:5, Eichler:2007:72, Oganessian:2012:162501, Sato:2015:209, Laatiaoui:2016:495, Chhetri:2018:263003, Raeder:2018:232503}, prompt the theory to study these complex many-particle systems. The interest is also fueled by the fact that a strong interplay between the correlation, relativistic, and quantum-electrodynamics (QED) effects for a large amount of core and valence electrons may manifest itself in qualitatively new properties of superheavy elements compared to their lighter homologues. In this regard, the ability of oganesson (Og, $Z\!=\!118$), despite its noble-gas electronic configuration, to form a negative ion has already become a textbook example~\cite{Eliav:1996:5350, Goidenko:2003:020102_R, Lackenby:2018:042512, Guo:2021:107, Kaygorodov:2021:012819}. 

The question of whether the periodic law holds for chemical elements beyond the seventh period~\cite{Fricke:1971:235, Seaborg:1996:3899, Nefedov:2006:149, Pyykko:2011:161, Jerabek:2018:053001, Kaygorodov:2020:036} is one of the fundamental motivations to study the physics and chemistry of superheavy elements. These investigations are unfeasible without the proper treatment of the QED effects. The state-of-the-art QED calculations of the middle- and high-$Z$ systems are performed within the $1/Z$ perturbation theory or its generalization based on a modification of the zeroth-order approximation by including some effective screening potential, see, e.g., Refs.~\cite{Sapirstein:2008:25,Glazov:2011:71,Volotka:2013:636,Shabaev:2018:60,Indelicato:2019:232001} for review. The corresponding sophisticated and laborious \textit{ab initio} methods cannot be directly incorporated into standard approaches based on the Dirac--Coulomb--Breit Hamiltonian~\cite{Grant:1970:747, Desclaux:1975:31, Bratzev:1977:173, Indelicato:1992:2426, Dzuba:1996:3948, Safronova:1999:4476, Tupitsyn:2003:022511, Kozlov:2015:199, Dzuba:2017:012503, Glazov:2017:46, Saue:2020:204104}. For this reason, there is a vital need for a simple and effective approximation for taking into account the QED corrections in the electronic-structure calculations. A number of such approaches has been proposed in the literature~\cite{Indelicato:1990:5139, Pyykko:2003:1469, Draganic:2003:183001, Flambaum:2005:052115, Thierfelder:2010:062503, Pyykko:2012:371, Tupitsyn:2013:682, Ginges:2016:052509}. For the purpose of describing the QED effects on binding and transition energies in relativistic many-electron systems, our group has suggested the model-QED operator in Ref.~\cite{Shabaev:2013:012513}. This operator has been successfully applied to the approximate QED calculations in various atomic systems including the superheavy ones~\cite{Tupitsyn:2016:253001, Pasteka:2017:023002, Yerokhin:2017:042505:2017:069901:join_pr, Machado:2018:032517, Si:2018:012504, Muller:2018:033416, Kaygorodov:2019:032505, Zaytsev:2019:052504, Shabaev:2020:052502, Kaygorodov:2021:012819, Savelyev:2022:012806, Kaygorodov:2022:preprint}, see also Refs.~\cite{Yerokhin:2020:042816, Skripnikov:2021:201101}.

\begin{figure}
\begin{center}
\includegraphics[width=0.75\columnwidth]{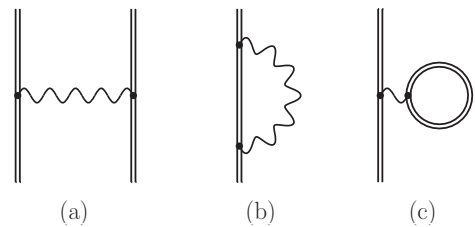}
\caption{\label{fig:qed_1order}
Lowest-order QED terms: one-photon-exchange~(a), self-energy~(b), and vacuum-polarization (c) diagrams. The double line corresponds to the electron propagator in the local binding potential. The wavy line denotes the photon propagator.}
\end{center}
\end{figure}

The model-QED-operator approach is worked out within the two-time Green's function (TTGF) method~\cite{TTGF} and based on the fact that the QED corrections can be systematically treated by constructing an effective Hamiltonian which acts in an appropriate active space~\cite{Shabaev:1993:4703}. In the case of \textit{ab initio} QED calculations for highly charged ions, unperturbed wave functions for a single level or a set of few quasi-degenerate states generally span the active space, see, e.g., Refs.~\cite{Yerokhin:2001:032109, Artemyev:2005:062104, Malyshev:2021:183001} and references therein. For the needs of the model-QED-operator construction, one should include into the active space all the Slater determinants made up of positive-energy Dirac-equation eigenfunctions with the total (many-electron) energies lying lower than the pair-creation threshold. Since our consideration is restricted to the lowest-order QED terms depicted in Fig.~\ref{fig:qed_1order}, the active space can actually be extended beyond this threshold~\cite{Shabaev:1993:4703}. The model-QED-operator approach formulates the effective Hamiltonian in a form suitable for the atomic-structure calculations. The \texttt{QEDMOD} Fortran package for computing the model-QED operator in the range $3 \leqslant Z \leqslant 120$ was presented in Ref.~\cite{Shabaev:2015:175:2018:69:join_pr}. The main goal of the present paper is to extend the region of supported nuclear charges up to $Z=170$. The nuclear charge $Z\approx 173$ is usually considered as a critical one, at which the $1s$ state of a hypothetical hydrogenlike ion with the extended nucleus reaches the negative-energy continuum~\cite{Pomeranchuk:1945:97, Gershtein:1970:358, Pieper:1969:327, Zeldovich:1972:673}. Therefore, the proposed operator should make it possible to study the QED effects on binding and transition energies in a wide range of the superheavy elements. Merging the model-QED operator with the various electron-correlation methods allows one to take the advantage of the rigorous calculations for hydrogenlike ions in many-electron systems where the \textit{ab initio} treatment is currently rather problematic.

In the next section, we briefly describe the structure of the model-QED operator and outline the modifications made compared to Ref.~\cite{Shabaev:2013:012513}. Then we discuss the evaluation of the self-energy and vacuum-polarization matrix elements within the rigorous QED approach in the range $110 \leqslant Z \leqslant 170$. Finally, the model-QED operator is tested by performing the radiative-correction calculations for superheavy ions with hydrogen- and alkali-metal-like electronic configurations as well as for neutral atoms and comparing the obtained results with the \textit{ab initio} ones.

Relativistic units ($\hbar=1$ and $c=1$) and Heaviside charge unit ($e^2=4\pi\alpha$, where $e<0$ is the electron charge) are used throughout the paper.


\section{Model-QED operator \label{sec:1}}

The Dirac equation represents the natural zeroth-order approximation for middle- and high-$Z$ atomic systems:
\begin{align}
\label{eq:dirac}
h^{\rm D} \psi \equiv 
\left[ \balpha \cdot \bp + \beta m + V \right] \psi = 
\veps \psi \, .
\end{align}
In the case of the potential induced by a point nucleus, $V(r)=-\alpha Z/r$, the eigenvalue of the Dirac Hamiltonian for the principal quantum number~$n$ and relativistic angular quantum number~$\kappa=(-1)^{j+l+1/2}(j+1/2)$ reads as
\begin{align}
\label{eq:en_point}
\veps_{n\kappa} = m \left[ 1 + \left( \frac{\alpha Z}{n_r + \lambda} \right)^2 \right]^{-1/2} \, ,
\end{align}
where $n_r=n-|\kappa|$ and $\lambda = \sqrt{\kappa^2-(\alpha Z)^2}$. The solutions for $|\kappa|=1$ do not formally exist for $Z>137$ due to a singularity at $\alpha Z=1$. In order to deal with the higher values of $Z$, one has to regularize the Hamiltonian~(\ref{eq:dirac}), see, e.g., Ref.~\cite{Gitman:2013:038104}. The most direct way is to employ a more realistic nuclear-potential model, i.e., to take into account the finite size of the nucleus. Then the energy of the $1s$ state will keep decreasing with $Z$ and at $Z=170$ almost reach the onset of the negative-energy continuum. Despite the fact that for high $Z$ the binding energy of low-lying levels exceeds $m$ and, accordingly, $\veps_{n\kappa}<0$, we will refer to these states as the positive-energy ones to distinguish them from the negative-energy continuum. Attributing the electron--nucleus interaction to the initial approximation with subsequent accounting for interactions of electrons with each other and with the quantized electromagnetic field by perturbation theory leads to the Furry picture of QED~\cite{Furry:1951:115}.

To the first order in $\alpha$, the QED effects can be described by an effective Hamiltonian acting in the subspace which is spanned by Slater determinants made up of the positive-energy solutions of Eq.~(\ref{eq:dirac}), see Ref.~\cite{Shabaev:1993:4703} for details. The total effective Hamiltonian can be expressed as 
\begin{align}
\label{eq:H_eff}
H = \Lambda^{(+)}
\left[ 
\sum_i \left( h_i^{\rm D} + h_i^{\rm QED} \right)
+ 
\sum_{i<j} h_{ij}^{\rm int}
\right]
\Lambda^{(+)}  \, ,
\end{align} 
where the sums run over all the atomic electrons and $\Lambda^{(+)}$ is the product of the one-electron projectors on the positive-energy eigenfunctions of $h^{\rm D}$. The interaction term~$h^{\rm int}$ arises from the one-photon exchange diagram in Fig.~\ref{fig:qed_1order}(a), see Refs.~\cite{Shabaev:1993:4703, TTGF, Shabaev:2013:012513}. We draw attention to the fact that the whole discussed formalism remains true if the potential~$V$ in Eq.~(\ref{eq:dirac}) includes some local screening potential $V_{\rm scr}$. In this case, the operator~$h^{\rm int}$ corresponds to residual electron--electron interaction. The term~$h^{\rm QED}$ represents the one-electron QED operator originating from the self-energy (SE) and vacuum-polarization (VP) diagrams in Figs.~\ref{fig:qed_1order}(b) and \ref{fig:qed_1order}(c), respectively. Within the TTGF method, one can derive the following symmetric expression for it~\cite{Shabaev:1993:4703}
\begin{align}
\label{eq:H_qed}
\!\!\!
h^{\rm QED} &\equiv  h^{\rm SE} + h^{\rm VP} =
\!\!\!\! \sum_{i,k}^{\veps_i,\veps_k>-m} \!\!\!\!
| \psi_i \rangle \langle \psi_i | 
\nonumber \\
& \times
\left\{
\frac{1}{2} \left[ \Sigma^{\rm SE}_R(\veps_i) + \Sigma^{\rm SE}_R(\veps_k) \right] + V^{\rm VP}_R
\right\}
| \psi_k \rangle \langle \psi_k | \, ,
\end{align} 
where $\Sigma^{\rm SE}_R(\veps)$ and $V^{\rm VP}_R$ are the renormalized SE and VP operators, respectively, and the sums over $i$ and $k$ run over all the positive-energy one-electron Dirac states. Below we discuss how the operator~$h^{\rm QED}$ can be adapted for convenient use in the practical relativistic electronic-structure calculations. In this section, we do not address the issue of the \textit{ab initio} calculations of the SE and VP corrections, see the next section for the discussion and relevant references.

Let us start with the SE part of the QED operator~(\ref{eq:H_qed}). We have to solve two issues simultaneously. On the one hand, due to the lack of simple enough algorithms for the \textit{ab initio} calculations of the $\Sigma$-operator matrix elements for arbitrary levels (including the continuum-spectrum states), one has to restrict the summation in Eq.~(\ref{eq:H_qed}) to a finite number of the low-lying one-electron eigenfunctions of~$h^{\rm D}$. On the other hand, it is necessary to ensure the short interaction range for the operator~$h^{\rm SE}$. In Ref.~\cite{Shabaev:2013:012513}, it was suggested to represent $h^{\rm SE}$ by a sum of short-range semilocal and nonlocal potentials. The nonlocal part was defined using the set of functions which are localized at smaller distances than the Dirac--Coulomb ones. The separation of the semilocal part with the support of the order of the Compton wavelength, $\lambdabar =1/m$, was justified by the fact that for low-$Z$ systems the dominant part of the SE correction indeed can be described by the local term~\cite{Welton:1948:1157}. For the range of the nuclear charges studied in the present work, we found, however, that the SE operator is hardly described with a simple local formula. For this reason, we now retain only the nonlocal potential in the model operator. So, the one-electron SE operator is approximated as follows
\begin{align}
\label{eq:se_model}
\tilde h^{\rm SE} = \sum_{j,l}^n 
| \phi_j \rangle B_{jl} \langle \phi_l | \, ,
\end{align}
where, as in Ref.~\cite{Shabaev:2013:012513}, we chose the functions $\{\phi_i\}_{i=1}^n$ to be
\begin{align}
\label{eq:proj}
\phi_i(\br) = \frac{1}{2} \left[ \, I - (-1)^{s_i} \beta \, \right] \rho_{l_i}(r) \psi_i(\br) \, . 
\end{align}
Here the index $s_i=n_i-l_i$ (with $n_i$ being the principal quantum number and $l_i=|\kappa_i+1/2|-1/2$ being the orbital angular momentum) enumerates the positive-energy states for the given angular symmetry, $I$ and $\beta$ are the identity and the standard Dirac matrices, respectively, and the factors $\rho_{l_i}(r) = \exp \left[ -2\alpha Z(r/\lambdabar)/(1+l_i) \right]$ serve to provide the stronger localization of the functions $\{\phi_i\}_{i=1}^n$ as compared to the Dirac--Coulomb ones $\{\psi_i\}_{i=1}^n$ (in practical calculations, we found that the replacement $1+l_i \rightarrow |\kappa_i|$ in $\rho_{l_i}$, affecting only the positive values of $\kappa_i$, may additionally improve the performance of the model-QED-operator approach  for $Z \gtrsim 160$). Finally, the coefficients $B_{jl}$ in Eq.~(\ref{eq:se_model}) are determined from the condition that the model operator~$\tilde h^{\rm SE}$ has to reproduce exactly the matrix elements of the operator~$h^{\rm SE}$ in the space spanned by the functions $\{\psi_i\}_{i=1}^n$, that is
\begin{align}
\label{eq:condition}
\sum_{j,l}^n 
\langle \psi_i | \phi_j \rangle B_{jl} \langle \phi_l | \psi_k \rangle
=
\frac{1}{2}
\langle \psi_i |
\left[ \Sigma^{\rm SE}_R(\veps_i) + \Sigma^{\rm SE}_R(\veps_k) \right]
| \psi_k \rangle
\end{align}
for $i,k=1\ldots n$, see Ref.~\cite{Shabaev:2013:012513} for details. We stress that the SE operator conserves the angular quantum numbers, and the matrix $B_{jl}$ has, accordingly, a block diagonal structure. 

The one-electron VP operator is equal to the sum of two local potentials, the Uehling and Wichmann--Kroll (WK) ones, $V^{\rm VP}_R = V_{\rm Ue} + V_{\rm WK}$. The dominant Uehling contribution is given by the expression
\begin{align}
\label{eq:Ue}
V_{\rm Ue}(r) = - &\frac{2\alpha^2 Z}{3mr} \, 
\int_0^\infty \! dr' \, r' \rho(r')    \nonumber \\
&\times \left[ K_0(2m|r-r'|) - K_0(2m|r+r'|) \right] \, , 
\end{align}
where
\begin{align}
\label{eq:K0}
K_0(x) = \int_1^\infty \! dt \,
e^{-xt} \left( \frac{1}{t^3} + \frac{2}{t^5} \right) \sqrt{t^2-1} \, ,
\end{align}
and the inducing-charge density~$\rho$ is normalized in accordance with $\int\!d{\br}\,\rho(r)=1$. The potential~$V_{\rm Ue}$ can be easily calculated either directly or using the approximate formulas from Ref.~\cite{Fullerton:1976:1283}. The evaluation of the WK part of the VP operator represents a much more difficult problem, see the discussion below. In Ref.~\cite{Shabaev:2013:012513} to a sufficient level of accuracy, this issue was solved by employing the approximate expressions obtained in Ref.~\cite{Fainshtein:1991:559}. These expressions were derived for the point-nucleus case that makes them unsuitable for superheavy elements. Therefore, in the present work we have performed the \textit{ab initio} calculations of the potential~$V_{\rm WK}$ for extended nuclei. For the sake of convenience, we approximate this contribution in a way similar to that applied for the operator~$h^{\rm SE}$. Namely, to represent the WK part of~$h^{\rm VP}$, the operator $\tilde{h}^{\rm WK}$ analogous to Eq.~(\ref{eq:se_model}) is introduced with the matrix~$B_{jl}$ determined by Eq.~(\ref{eq:condition}), where the evaluated WK potential stand instead of the SE operator, $\Sigma^{\rm SE}_R \rightarrow V_{\rm WK}$. The Uehling term is treated as in Ref.~\cite{Shabaev:2013:012513} without any changes. In principle, the SE operator and the WK part of the VP potential can be considered together during the procedure of the model-QED-operator construction. Then the nonlocal potential~$\tilde{h}^{\rm SE+WK}$ with the appropriately defined coefficients $B_{jl}$ will simultaneously approximate the SE and WK contributions.

Concluding the description of the model-QED operator, it should be noted that in what follows, as in Ref.~\cite{Shabaev:2013:012513}, we construct the model operator employing the functions~(\ref{eq:proj}) which correspond to the $ns$ states with the principal quantum number $n \leqslant 3$ and the $np_{1/2}$, $np_{3/2}$, $nd_{3/2}$, and $nd_{5/2}$ states with $n \leqslant 4$.


\section{Rigorous QED evaluation of the SE and VP matrix elements \label{sec:2}}

The \textit{ab initio} calculations of the SE and VP matrix elements are performed for extended nuclei employing the standard two-parameter Fermi model for the nuclear-charge distribution. The size of the nuclei is determined using the approximate formulas given in Ref.~\cite{Pieper:1969:327}. First, we relate the atomic mass number $A$ with the charge $Z$ via
\begin{align}
\label{eq:A}
A = 0.00733 \, Z^2 + 1.30 \, Z + 63.6 \, ,
\end{align}
rounding it up to the nearest integer. Second, we evaluate the root-mean-square radius~$R$ in fm according to 
\begin{align}
\label{eq:R}
R=\sqrt{\frac{3}{5}} \, R_{\rm sphere} \, ,
\end{align}
where $R_{\rm sphere} = 1.2\, A^{1/3}$. The uncertainties of the presented below results do not include errors associated with the choice of $R$ and the nuclear model, but are determined only by studying numerical aspects.

\begin{figure}
\begin{center}
\includegraphics[width=0.94\columnwidth]{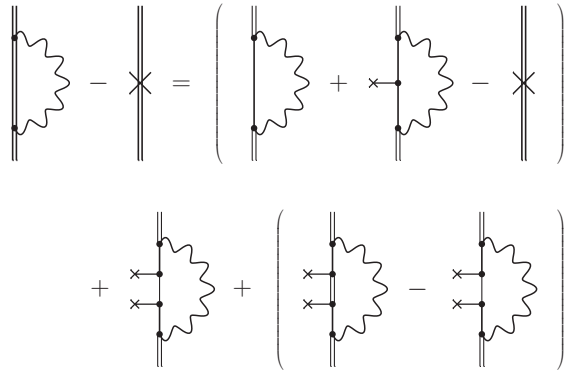}
\caption{\label{fig:se}
Decomposition of the self-energy diagram. The single line represents the free-electron propagator. The double line with a cross on it corresponds to the mass counterterm. The line ended with a small cross denotes the interaction with the binding potential~$V$.}
\end{center}
\end{figure}

\begin{figure}
\begin{center}
\includegraphics[width=\columnwidth]{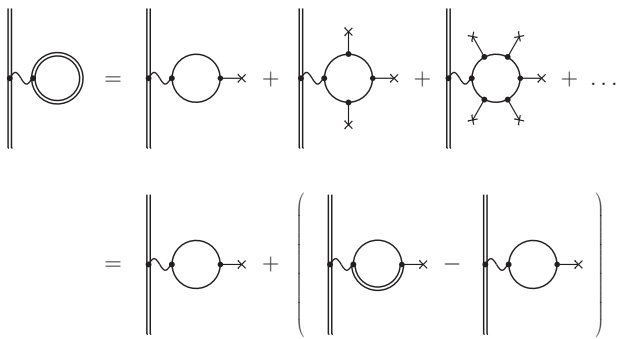}
\caption{\label{fig:vp}
Decomposition of the vacuum-polarization diagram.}
\end{center}
\end{figure}

The one-loop SE operator reads as follows, see, e.g., Ref.~\cite{Mohr:1998:227},
\begin{align}
\label{eq:Sigma}
\Sigma_R^{\rm SE}(\veps,\br_1,\br_2) &= 
2 i \alpha \int_{-\infty}^\infty \!\! d\omega \,
D^{\mu\nu}(\omega,\br_{12})  \nonumber \\
&\times
\alpha_\mu G(\veps-\omega,\br_1,\br_2)\alpha_\nu  - \beta \delta m \, ,
\end{align}
where $\delta m$ is the mass counterterm, $D^{\mu\nu}$ is the photon propagator, $G$ is the Dirac--Coulomb Green's function (compared to Ref.~\cite{Mohr:1998:227}, we define it with the opposite sign, $G(\omega,\br_1,\br_2)=(\omega-h^{\rm D})^{-1}$), $\alpha^\mu = \gamma^0 \gamma^\mu$ and $\beta = \gamma^0$ are the Dirac matrices, and it is implicitly assumed that the integration contour extending from $-\infty$ to $\infty$ provides the proper bypass of all the singularities in the complex $\omega$-plane. The nonperturbative (in $\alpha Z$) calculations of the SE contribution have been extensively discussed in the literature, for review see Refs.~\cite{Mohr:1974:26:1974:52:join_pr, Snyderman:1991:43, Blundell:1991:R1427, Mohr:1993:158, Jentschura:1999:53, Cheng:1993:1817, Yerokhin:1999:800} and references therein. The expression~(\ref{eq:Sigma}) suffers from the ultraviolet divergences. To isolate them and calculate the diagonal and off-diagonal matrix elements of the resulting renormalized operator $\Sigma^{\rm SE}_R$ with the Dirac--Coulomb wave functions, we employ the method worked out in Ref.~\cite{Yerokhin:1999:800} with some modifications proposed in Refs.~\cite{Artemyev:2007:173004, Artemyev:2013:032518}. The method is based on the expansion of the bound-electron propagator in powers of the interaction with the binding potential~$V$. The divergences are contained only in the so-called zero- and one-potential contributions shown in brackets in the first line of Fig.~\ref{fig:se}. These two terms are renormalized together with the mass counterterm and evaluated in the momentum representation. In principle, the remaining part of the SE operator is finite and can be calculated in the coordinate space. However, aiming to improve the partial-wave-expansion convergence, we additionally separate the slowly convergent term with two interactions. The corresponding two-potential contribution is evaluated using the analytical expression for the free-electron propagator~\cite{Mohr:1974:26:1974:52:join_pr}. The higher-order remainder shown in brackets in the second line of Fig.~\ref{fig:se} is calculated as the pointwise difference of the two similar expressions in the $\omega$-integration by employing the finite-basis-set representation for the free- and bound-electron Green's functions. We will refer to this term as the many-potential one. We note that in Ref.~\cite{Yerokhin:1999:800} such a designation was applied to the contribution with two and higher interactions. The finite basis sets are constructed from B splines~\cite{Johnson:1988:307, Sapirstein:1996:5213} in the framework of the dual-kinetic-balance approach~\cite{splines:DKB}. In the calculations, the maximal value of $|\kappa|$ reaches 45 and 20 for the two- and many-potential terms, respectively. The remainders of the $|\kappa|$-series are estimated using a polynomial fitting in $1/|\kappa|$.

The vacuum-polarization potential can be written as follows, see, e.g., Ref.~\cite{Mohr:1998:227},
\begin{align}
\label{eq:V_vp}
V^{\rm VP} (\br) = \frac{\alpha}{2\pi i} 
\int \! d\br' 
\int_{-\infty}^\infty \!\! d\omega \,
\frac{{\rm Tr} \, G(\omega,\br',\br')}{|\br-\br'|} \, .
\end{align} 
The formal expression~(\ref{eq:V_vp}) is ultraviolet divergent, and it demands the charge renormalization. To do so, one has to expand the Dirac--Coulomb Green's function~$G$ in terms of the free-electron propagators. According to the Furry's theorem~\cite{Furry:1937:125}, only the fermion loops with an even number of vertices contribute, whereas the diagrams with an odd number of the interactions vanish owing to a charge conjugation argument, see the first line of Fig.~\ref{fig:vp}. Thereby, the first nonzero contribution arises from the diagram linear with respect to the external field~$V$, being of order $\alpha(\alpha Z)$. Applying the charge-renormalization procedure to this term, one arrives at the Uehling potential given by Eq.~(\ref{eq:Ue}), see Refs.~\cite{Uehling:1935:55, Serber:1935:49}. The remaining part of the vacuum-polarization potential, which includes the terms of order $\alpha (\alpha Z)^3$ and higher, corresponds to the WK contribution. The all-order (in $\alpha Z$) calculations of this potential have a long history, for review see, e.g., Refs.~\cite{Wichmann:1956:843, Brown:1975:581:1975:596:1975:609:join_pr, Gyulassy:1975:497, Soff:1988:5066, Manakov:1989:673, Persson:1993:2772, Sapirstein:2003:042111} and references therein. The WK potential is finite. However, a special care has to be taken when dealing with a light-by-light scattering contribution represented by the second term of the expansion in Fig.~\ref{fig:vp}, since it may contain a spurious gauge-noninvariant piece. Without going into details, we just point out that a single subtraction of the Uehling contribution from the total vacuum-polarization potential leads to a correct result for the WK potential provided the calculations are arranged so that the terms with positive and negative values of $\kappa$ in the electron propagator are treated together~\cite{Gyulassy:1975:497}, see also Ref.~\cite{Soff:1988:5066}. This prescription for the evaluation of the WK potential is shown schematically in brackets in the second line of Fig.~\ref{fig:vp}, where we additionally removed the free-electron-propagator contribution from the Dirac--Coulomb Green's function which is zero due to the Furry's theorem. It is convenient to rotate the integration contour in the complex $\omega$-plane to the imaginary axis and change the variable of integration to $\eta$, where $\omega=i\eta$. During this rotation, the bound-state poles of the Green's function~$G$ located on the negative real $\omega$-axis are picked up as residues. Finally, taking into account that the induced vacuum-polarization charge must be zero, we can express the WK potential as follows
\begin{align}
\label{eq:WK}
V_{\rm WK}(r) &= \frac{2\alpha}{\pi} \,
\int_r^\infty \! dr' \, r' \left( 1 - \frac{r'}{r} \right)   \nonumber \\
&\times
\left[\, 
\sum_\kappa |\kappa| \, {\rm Re} \int_0^\infty \!\! d\eta \,\,
 {\rm Tr} \, G_\kappa^{(2+)}(i\eta,r',r') \right. \nonumber \\
&\quad\left.
- \,
\pi \! \sum^{-m<\veps_{n\kappa}<0}_{n\kappa} \! |\kappa| \left( g_{n\kappa}^2(r') + f_{n\kappa}^2(r') \right)
\,\right] \, ,
\end{align}
where 
\begin{align}
\label{eq:G2plus}
G_\kappa^{(2+)}(\omega,x,y) = 
&\int_0^\infty \! dz \, z^2 G^{(0)}_\kappa(\omega,x,z) V(z)   \nonumber \\
&\times
\left[ G_\kappa(\omega,z,y) - G^{(0)}_\kappa(\omega,z,y) \right]
\end{align}
with $G^{(0)}_\kappa$ and $G_\kappa$ being the radial parts of the free- and bound-electron Green's functions~\cite{Mohr:1998:227}, respectively. We find them numerically by solving a corresponding system of differential equations. In Eq.~(\ref{eq:WK}), $g_{n\kappa}$ and $f_{n\kappa}$ denote the large and small radial components of the Dirac wave function, normalized according to
\begin{align}
\label{eq:wf_norm}
\int_0^\infty \! dr \, r^2 \left( g_{n\kappa}^2(r) + f_{n\kappa}^2(r) \right) = 1 \, .
\end{align}
We terminate the summation over~$\kappa$ in Eq.~(\ref{eq:WK}) at $|\kappa|=9$--11 and study the convergence of the matrix elements $\langle \psi_i | V_{\rm WK} | \psi_k \rangle$, which are employed thereafter to determine the model-QED operator. In principle, to represent the evaluated WK potential within the model-QED-operator approach, besides the method suggested in the previous section, there is also an option to store it point-by-point on an appropriate integration grid.

\input{table_SE_s.tex}

\input{table_SE_p1.tex}

\input{table_SE_p3.tex}

\input{table_SE_d3.tex}

\input{table_SE_d5.tex}

\input{table_SE_170.tex}

The results of the \textit{ab initio} calculations of the first-order one-electron QED contributions are conveniently expressed in terms of the function~$F_{n_in_k}(\alpha Z)$ defined by
\begin{align}
\label{eq:F}
\langle \psi_i | h^{\rm QED} | \psi_k \rangle = 
\frac{\alpha}{\pi} \frac{(\alpha Z)^4}{(n_in_k)^{3/2}} \, F_{n_in_k}(\alpha Z) \, mc^2 \, ,
\end{align}
where $n_i$ and $n_k$ are the principal quantum numbers of the $i$ and $k$ states, respectively. Our results for the SE matrix elements obtained for the $ns$, $np_{1/2}$, $np_{3/2}$, $nd_{3/2}$, and $nd_{5/2}$ states with $n$ up to 5 are given in Tables~\ref{tab:se_s}, \ref{tab:se_p1}, \ref{tab:se_p3}, \ref{tab:se_d3}, and \ref{tab:se_d5}, respectively. As noted above, all the values are calculated for the extended nuclei. The uncertainties are obtained by studying the convergence with respect to the partial-wave expansion in the two- and many-potential terms as well as the dependence of the many-potential terms on the size of the finite basis set employed. If no error is specified, the value is assumed to be accurate to all digits quoted. For $Z=170$, the breakdown of the SE correction into the zero-, \mbox{one-,} two-, and many-potential terms is presented in Table~\ref{tab:se_170} for the lowest-energy levels with different values of $\kappa$. Our result (in terms of the function $F$) for the $1s$ state, $3.8831$, is in reasonable agreement with the value of $3.909$ given in Ref.~\cite{Soff:1982:1465}, where the nuclear size was adjusted in such a way that the $K$-electron energy differs only by 1~meV from the borderline of the negative-energy continuum and the homogeneously-charged sphere was assumed to describe the nuclear-charge distribution. In other cases in Ref.~\cite{Soff:1982:1465}, the atomic mass number was chosen to be $A=2.5Z$ that led to $F$ equal to $1.972$, $2.913$, and $3.517$ for $Z=130$, $Z=150$, and $Z=160$, respectively. These results are in agreement with our values of $1.9832$, $2.8941$, and $3.4565$ as well. Finally, we note a drastic growth of the SE contribution for the $np_{1/2}$ states in the high-$Z$ region, see Table~\ref{tab:se_p1}. For $Z=170$, the absolute value of the SE correction for the $2p_{1/2}$ state almost reaches in magnitude the corresponding value for the $1s$ state. Apparently, this trend is explained by the behavior of the small components of the wave function for the $np_{1/2}$ states which penetrate the region $r<\lambdabar$ for very large~$Z$.

\input{table_WK_s.tex}

\input{table_WK_p1.tex}

\input{table_WK_p3.tex}

\input{table_WK_d3.tex}

\input{table_WK_d5.tex}

The results for the matrix elements of the WK potential evaluated with the Dirac--Coulomb wave functions for the $ns$, $np_{1/2}$, $np_{3/2}$, $nd_{3/2}$, and $nd_{5/2}$ states with $n$ up to 5 are presented in Tables~\ref{tab:wk_s}, \ref{tab:wk_p1}, \ref{tab:wk_p3}, \ref{tab:wk_d3}, and \ref{tab:wk_d5}, respectively. For the $d$ states, the WK correction is small. Therefore, for a better representation of this contribution as a function of $Z$, in Tables~\ref{tab:wk_d3} and \ref{tab:wk_d5} we give an additional significant digit, compared to the other similar tables. In Ref.~\cite{Soff:1988:5066}, the extended nucleus was modeled by a homogeneously-charged spherical shell. For $Z=170$, the radius of the nucleus was assumed to be $R_{\rm shell} = 7.1$~fm and the WK contributions (in terms of the function $F$) for the $1s$, $2s$, and $2p_{1/2}$ states were found to be $0.519$, $0.766$, and $3.76$, respectively. These values are in reasonable agreement with our results given in Tables~\ref{tab:wk_s} and \ref{tab:wk_p1}. The qualitative agreement is found for the partial-wave contributions to the vacuum-polarization charge density as well. 

The results given in Tables~\ref{tab:se_s}--\ref{tab:se_d5} and \ref{tab:wk_s}--\ref{tab:wk_d5} are employed to represent the SE and WK contributions by means of the nonlocal model-QED operator in the range $110 \leqslant Z \leqslant 170$ according to the prescriptions formulated in Sec.~\ref{sec:1}. To obtain the function~$F_{n_in_k}(\alpha Z)$ for values of $Z$ not listed in the tables, a polynomial interpolation can be used
\begin{align}
\label{eq:interpol}
F_{n_in_k}(\alpha Z) = 
\sum_{n=1}^N F_{n_in_k}(\alpha Z_n) 
\prod_{m\neq n} \frac{Z-Z_m}{Z_n-Z_m} \, . 
\end{align} 
In contrast to Ref.~\cite{Shabaev:2013:012513}, we do not follow the receipt from Ref.~\cite{Mohr:1983:453}, which for the $s$ states implies the interpolation of the function $F_{n_in_k}(\alpha Z)$ with subtraction of the term describing the small-$\alpha Z$ behavior. For the range of $Z$ under consideration, this subtraction is not justified. In the overlapping region $110 \leqslant Z \leqslant 120$, both implementations of the model-QED-operator approach, the original~\cite{Shabaev:2013:012513} and the current one, are applicable and can be compared. We note, however, that in Ref.~\cite{Shabaev:2013:012513} the finite-nuclear-size corrections to the SE contributions were evaluated for slightly different values of nuclear radii. For this reason, in what follows, when comparing these two versions of the model-QED operator, we always use the matrix elements of the SE operator obtained in the present work. Therefore, the corresponding comparison boils down to the analysis of the operators with and without the local potential isolated.


\section{Test of the model-QED operator \label{sec:3}}

The model-QED operator is constructed employing the SE and WK matrix elements evaluated with the Dirac--Coulomb wave functions for the $ns$ states with $n \leqslant 3$ and the $np$ and $nd$ states with $n \leqslant 4$. So, by construction, the radiative corrections for these states are reproduced exactly with the operator. Therefore, the natural test is to probe the ``predictive'' power of the developed operator by calculating the QED corrections for the states with the higher values of the principal quantum number. In Tables~\ref{tab:se_s} and \ref{tab:wk_s}, the matrix elements for the $4s$ state are given. We have performed the \textit{ab initio} calculations of the diagonal matrix elements for the $n=5$ states for a number of $Z$ as well. In Table~\ref{tab:se_prediction}, we present the SE contributions for the $4s$, $5s$, $5p_{1/2}$, $5p_{3/2}$, $5d_{3/2}$, and $5d_{5/2}$ states in hydrogenlike ions. The rows labeled with ``Exact'' correspond to the results of the \textit{ab initio} calculations taken from Tables~\ref{tab:se_s}--\ref{tab:se_d5}, whereas the values in the lines ``Mod. op.'' and ``Ref.~\cite{Shabaev:2013:012513}'' are obtained by averaging the current and original versions of the model-QED operator~$\tilde{h}^{\rm SE}$ with the Dirac--Coulomb wave functions, respectively. For the $s$ states, the deviation of the model-QED-operator predictions from the exact values does not exceed 1\%. For $5d_{3/2}$, the situation is slightly worse than for the other states. It is explained mainly by the smallness of the SE correction, since for all the $nd_{3/2}$ states the corresponding functions~$F_{n_in_k}$ change the sign at $Z\approx 110$. In general, the model-QED operator leads to the values for the SE contributions which are very close to the \textit{ab initio} ones.

In Table~\ref{tab:wk_prediction}, we give the similar comparison for the WK contribution. The notations are the same as in Table~\ref{tab:se_prediction}. Again, for demonstration purposes we show extra digits for the function~$F$ in the cases of the $5p_{3/2}$ and $5d$ states because of the smallness of these contributions. From Table~\ref{tab:wk_prediction} it is seen, that the proposed nonlocal potential reproduces the WK contribution to good accuracy. 

In Table~\ref{tab:se_alkali}, we demonstrate the capacity of the model-QED-operator approach to the radiative-correction calculations in many-electron ions of superheavy elements possessing alkali-metal-like configurations, namely, ${\rm [Ne]}3s$, ${\rm [Ne]}3s^2 3p^6 4s$, and ${\rm [Ne]}3s^2 3p^6 3d^{10} 4s^2 4p^6 5s$. First, following Refs.~\cite{Labzowsky:1999:2707, Sapirstein:2002:042501}, we have evaluated the one-loop SE contributions for the valence $ns$ electrons within the rigorous QED approach including a local screening potential into the initial approximation, see the related discussion after Eq.~(\ref{eq:H_eff}) and, e.g., Refs.~\cite{TTGF, Sapirstein:2008:25} for review. We have chosen the Kohn--Sham potential~\cite{pot:KS} with the Latter correction~\cite{Latter:1955:510} introduced for restoring the proper asymptotic behavior to model the interelectronic-interaction effects in the zeroth-order Hamiltonian. The results of these \textit{ab initio} calculations are given in rows labeled ``Exact''. Then the model-QED-operator predictions have been obtained by averaging the operator~$\tilde{h}^{\rm SE}$ with the valence-electron wave functions determined from the Dirac equation~(\ref{eq:dirac}), in which the nuclear potential is replaced with the Kohn--Sham one. As in Table~\ref{tab:se_prediction}, for $Z=110$ and $Z=120$ we present also the results evaluated by means of the original version of the operator~\cite{Shabaev:2013:012513}. For completeness, the \textit{ab initio} values from Table~\ref{tab:se_s} corresponding to hydrogenlike ions are shown in lines labeled ``H-like''. From Table~\ref{tab:se_alkali}, it can be seen that the model-QED operator constructed using the rigorous calculations with the Dirac--Coulomb basis works reasonably well in this nonhydrogenic case.

In Ref.~\cite{Cheng:1976:1943}, the SE shifts for the $1s$ level in superheavy elements were evaluated employing the Dirac--Fock--Slater potential constructed with the Slater-exchange term~\cite{Slater:1951:385}. In Table~\ref{tab:se_1s_DHFS}, we compare the results of Ref.~\cite{Cheng:1976:1943} with our theoretical predictions. The lines labeled ``Mod. op.'', ``Ref.~\cite{Shabaev:2013:012513}'', and ``H-like'' have the same meaning as in Table~\ref{tab:se_alkali}. We have generated the Dirac--Fock--Slater potential for neutral atoms of superheavy elements in accordance with the relativistic ground-state configurations given in Ref.~\cite{Fricke:1977:83}. From Table~\ref{tab:se_1s_DHFS}, it is seen how the averaging of the model-QED-operator with the corresponding wave functions shifts the SE corrections in such a way that they are in perfect agreement with the values from Ref.~\cite{Cheng:1976:1943} within the indicated error bars.

In addition, we mention that averaging the model-QED operator with one- or many-electron wave functions does not exhaust all the possibilities for applying this approach. As noted in Ref.~\cite{Shabaev:2013:012513}, this operator can be self-consistently included into the Dirac--Fock equations, also referred to as the relativistic Hartree--Fock ones. This may be especially important in cases when the leading-order contributions cancel out for some reasons, see, e.g., Ref.~\cite{Shabaev:2020:052502}. Treating the model-QED operator in this way allows one to take into account single-particle excitations into the negative-energy continuum as well as to partly include the higher-order QED contributions.

\input{table_SE_prediction.tex}

\input{table_WK_prediction.tex}

\input{table_SE_alkali.tex}

\input{table_SE_1s_DHFS.tex}


\section{Conclusions \label{sec:4}}

The model-QED-operator approach, proposed in Ref.~\cite{Shabaev:2013:012513}, has been extended to the region of nuclear charges $110 \leqslant Z \leqslant 170$. The self-energy and the Wichmann--Kroll part of the vacuum-polarization potential are represented by nonlocal operators. The model-QED operator can be easily incorporated into any of the existing methods for solving the Dirac--Coulomb--Breit equation. The capacity of the approach has been demonstrated for a number of systems by comparing the model-QED-operator predictions with the results of the corresponding \textit{ab initio} calculations. The developed model-QED operator can be used to evaluate the QED effects in superheavy elements in a wide range of~$Z$.


\section*{Acknowledgments}

The work is supported by the Ministry of Science and Higher Education of the Russian Federation within the Grant No. 075-10-2020-117.




\end{document}

%% file: table_SE_s.tex
\begin{table*}[t]
\centering

\renewcommand{\arraystretch}{1.1}

\caption{\label{tab:se_s} 
         Matrix elements of the self-energy operator for the $ns$ states evaluated with the Dirac--Coulomb wave functions.
         Labels $(n_i,n_k)$ stand for the function~$F_{n_in_k}$ defined by Eq.~(\ref{eq:F}).
         $R$ is the root-mean-square charge radius of the nuclei (in fm).
         }
         
\resizebox{\textwidth}{!}{%
\begin{tabular}{
                S[table-format=3.0]@{\quad}
                S[table-format=2.4]
                S[table-format=2.5]@{\!}
                S[table-format=2.5]@{\!}
                S[table-format=2.5]@{\!}
                S[table-format=2.5]@{\!}
                S[table-format=2.5]@{\!}
                S[table-format=2.5]@{\!}
                S[table-format=2.5]@{\!}
                S[table-format=2.5]@{\!}
                S[table-format=2.5]@{\!}
                S[table-format=2.4(1)]
                S[table-format=2.4(1)]
                S[table-format=2.5]@{\!}
                S[table-format=2.4(1)]
                S[table-format=2.4(1)]
                S[table-format=2.4(1)]@{}
               }
               
\hline
\hline

   \multicolumn{1}{c}{\rule{0pt}{1.2em}$Z$~~~}      &
   \multicolumn{1}{c}{$R$~~}                        &
   \multicolumn{1}{c}{(1,1)}                        &
   \multicolumn{1}{c}{(1,2)}                        &
   \multicolumn{1}{c}{(1,3)}                        &
   \multicolumn{1}{c}{(1,4)}                        &
   \multicolumn{1}{c}{(1,5)}                        &
   \multicolumn{1}{c}{(2,2)}                        &
   \multicolumn{1}{c}{(2,3)}                        &
   \multicolumn{1}{c}{(2,4)}                        &
   \multicolumn{1}{c}{(2,5)}                        &
   \multicolumn{1}{c}{(3,3)}                        &
   \multicolumn{1}{c}{(3,4)}                        &
   \multicolumn{1}{c}{(3,5)}                        &
   \multicolumn{1}{c}{(4,4)}                        &
   \multicolumn{1}{c}{(4,5)}                        &
   \multicolumn{1}{c}{(5,5)}                        \\      
        
\hline   
                       
 110 & 6.188 &      1.5744 &      1.9995 &      1.9788 &      1.9353 &      1.9006 &      2.6283 &      2.6172 &      2.5695 &      2.5278 &      2.6009 &      2.5505 &      2.5104 &   2.4980(1) &   2.4559(1) &   2.4137(5)  \rule{0pt}{3.6ex}   \\ 
 115 & 6.291 &      1.6401 &      2.1172 &      2.0862 &      2.0322 &      1.9901 &      2.8439 &      2.8170 &      2.7535 &      2.7008 &      2.7882 &      2.7215 &      2.6704 &   2.6547(1) &   2.6015(1) &   2.5492(4)   \\ 
 120 & 6.391 &      1.7275 &      2.2666 &      2.2210 &      2.1532 &      2.1015 &      3.1127 &      3.0636 &      2.9790 &      2.9117 &      3.0174 &      2.9296 &      2.8639 &   2.8440(1) &   2.7769(1) &   2.7117(4)   \\ 
 125 & 6.494 &      1.8402 &      2.4519 &      2.3854 &      2.2994 &      2.2354 &      3.4427 &      3.3621 &      3.2490 &      3.1625 &      3.2917 &      3.1762 &      3.0916 &   3.0663(1) &   2.9815(1) &   2.9001(3)   \\ 
 130 & 6.588 &      1.9832 &      2.6774 &      2.5809 &      2.4709 &      2.3911 &      3.8397 &      3.7149 &      3.5634 &      3.4518 &      3.6109 &      3.4597 &      3.3507 &      3.3184 &   3.2116(1) &   3.1099(3)   \\ 
 135 & 6.691 &      2.1596 &      2.9412 &      2.8026 &      2.6615 &      2.5619 &      4.2976 &      4.1118 &      3.9099 &      3.7663 &      3.9627 &   3.7663(1) &      3.6268 &      3.5859 &   3.4525(1) &   3.3262(2)   \\ 
 140 & 6.785 &      2.3728 &      3.2372 &      3.0417 &      2.8616 &      2.7380 &      4.7978 &      4.5315 &      4.2662 &      4.0835 &      4.3250 &   4.0743(1) &      3.8982 &      3.8474 &   3.6832(1) &   3.5285(2)   \\ 
 145 & 6.883 &      2.6203 &      3.5445 &      3.2777 &      3.0518 &      2.9007 &      5.2928 &      4.9293 &      4.5900 &      4.3628 &   4.6570(1) &   4.3459(1) &      4.1293 &   4.0680(1) &   3.8711(1) &   3.6860(2)   \\ 
 150 & 6.983 &      2.8941 &      3.8316 &      3.4851 &      3.2099 &      3.0298 &      5.7178 &      5.2514 &      4.8341 &      4.5612 &   4.9159(1) &   4.5442(1) &      4.2866 &   4.2160(1) &   3.9875(1) &   3.7732(1)   \\ 
 155 & 7.075 &      3.1808 &      4.0637 &      3.6414 &      3.3182 &      3.1102 &      6.0137 &      5.4569 &      4.9665 &      4.6514 &   5.0765(1) &   4.6504(1) &      4.3556 &   4.2785(1) &   4.0227(1) &   3.7830(1)   \\ 
 160 & 7.170 &      3.4565 &      4.2051 &      3.7273 &      3.3632 &      3.1312 &      6.1433 &      5.5273 &      4.9772 &      4.6278 &   5.1351(1) &   4.6664(1) &      4.3411 &   4.2614(1) &   3.9842(1) &   3.7247(1)   \\ 
 165 & 7.268 &      3.6964 &      4.2393 &      3.7385 &      3.3445 &      3.0944 &      6.1212 &      5.4837 &      4.8897 &      4.5151 &   5.1181(1) &   4.6186(1) &      4.2697 &   4.1913(1) &   3.8982(1) &   3.6243(1)   \\ 
 170 & 7.358 &      3.8831 &      4.1722 &      3.6850 &      3.2731 &      3.0108 &      6.0026 &      5.3717 &      4.7466 &      4.3539 &      5.0656 &      4.5433 &      4.1759 &   4.1013(1) &   3.7956(1) &   3.5106(1)   \\ 

\hline
\hline

\end{tabular}%
}

\end{table*}

%% file: table_SE_p1.tex
\begin{table*}[t]
\centering

\renewcommand{\arraystretch}{1.1}

\caption{\label{tab:se_p1} 
         Matrix elements of the self-energy operator for the $np_{1/2}$ states evaluated with the Dirac--Coulomb wave functions.
         The notations are the same as in Table~\ref{tab:se_s}.
         }
         
\resizebox{\textwidth}{!}{%
\begin{tabular}{
                S[table-format=3.0]@{\quad}
                S[table-format=2.4]
                S[table-format=3.4(1)]
                S[table-format=3.4(1)]
                S[table-format=2.4(1)]
                S[table-format=2.4(1)]
                S[table-format=2.4(1)]
                S[table-format=2.4(1)]
                S[table-format=2.5]
                S[table-format=2.4(1)]
                S[table-format=2.4(1)]
                S[table-format=2.4(1)]@{}
               }
               
\hline
\hline

   \multicolumn{1}{c}{\rule{0pt}{1.2em}$Z$~~~}      &
   \multicolumn{1}{c}{$R$~~}                        &
   \multicolumn{1}{c}{(2,2)}                        &
   \multicolumn{1}{c}{(2,3)}                        &
   \multicolumn{1}{c}{(2,4)}                        &
   \multicolumn{1}{c}{(2,5)}                        &
   \multicolumn{1}{c}{(3,3)}                        &
   \multicolumn{1}{c}{(3,4)}                        &
   \multicolumn{1}{c}{(3,5)}                        &
   \multicolumn{1}{c}{(4,4)}                        &
   \multicolumn{1}{c}{(4,5)}                        &
   \multicolumn{1}{c}{(5,5)}                        \\      
        
\hline   
                       
 110 & 6.188 &      0.6952 &      0.7680 &      0.7754 &      0.7709 &      0.7992 &      0.8054 &      0.8033 &   0.7979(1) &   0.7925(1) &   0.7826(5)  \rule{0pt}{3.6ex}   \\ 
 115 & 6.291 &      0.8863 &      0.9638 &      0.9663 &      0.9569 &      0.9963 &      0.9957 &      0.9883 &   0.9815(1) &   0.9699(1) &   0.9543(4)   \\ 
 120 & 6.391 &      1.1524 &      1.2315 &      1.2243 &      1.2066 &      1.2618 &      1.2493 &      1.2330 &   1.2237(1) &   1.2025(1) &   1.1783(3)   \\ 
 125 & 6.494 &      1.5350 &      1.6071 &      1.5814 &      1.5493 &      1.6261 &      1.5928 &      1.5617 &      1.5478 &   1.5114(1) &   1.4733(3)   \\ 
 130 & 6.588 &      2.1038 &      2.1459 &      2.0844 &      2.0274 &      2.1316 &      2.0612 &      2.0050 &      1.9821 &   1.9213(1) &   1.8608(2)   \\ 
 135 & 6.691 &      2.9684 &      2.9225 &      2.7925 &      2.6915 &      2.8219 &   2.6863(1) &      2.5881 &      2.5478 &   2.4481(1) &   2.3519(2)   \\ 
 140 & 6.785 &      4.2916 &      4.0183 &      3.7616 &      3.5859 &   3.7155(1) &   3.4695(1) &      3.3051 &   3.2327(1) &   3.0745(1) &   2.9244(2)   \\ 
 145 & 6.883 &   6.2542(1) &   5.4550(1) &      4.9843 &      4.6919 &   4.7322(1) &   4.3216(1) &      4.0651 &   3.9410(1) &   3.7056(1) &   3.4843(1)   \\ 
 150 & 6.983 &   8.9471(1) &   7.1040(1) &   6.3267(1) &      5.8780 &   5.6582(1) &   5.0499(1) &      4.6895 &   4.5005(1) &   4.1833(1) &   3.8861(1)   \\ 
 155 & 7.075 &  12.2304(1) &   8.6895(1) &   7.5567(1) &      6.9345 &   6.2675(1) &   5.4810(1) &      5.0302 &   4.7814(1) &   4.3985(1) &   4.0394(1)   \\ 
 160 & 7.170 &  15.6906(2) &   9.9403(1) &   8.4719(1) &   7.6899(1) &   6.5099(1) &   5.6016(1) &      5.0894 &   4.7995(1) &   4.3764(1) &   3.9783(1)   \\ 
 165 & 7.268 &  18.8730(2) &  10.7733(1) &   9.0255(1) &   8.1135(1) &   6.5299(1) &   5.5454(1) &      4.9957 &   4.6797(1) &   4.2347(1) &   3.8144(1)   \\ 
 170 & 7.358 &  21.5128(2) &  11.2711(1) &   9.2925(1) &   8.2788(1) &   6.5078(1) &      5.4573 &      4.8792 &   4.5417(1) &   4.0797(1) &   3.6434(1)   \\ 

\hline
\hline

\end{tabular}%
}

\end{table*}

%% file: table_SE_p3.tex
\begin{table*}[t]
\centering

\renewcommand{\arraystretch}{1.1}

\caption{\label{tab:se_p3} 
         Matrix elements of the self-energy operator for the $np_{3/2}$ states evaluated with the Dirac--Coulomb wave functions.
         The notations are the same as in Table~\ref{tab:se_s}.
         }
         
\begin{tabular}{
                S[table-format=3.0]@{\quad}
                S[table-format=2.4]
                S[table-format=2.5]
                S[table-format=2.5]
                S[table-format=2.5]
                S[table-format=2.5]
                S[table-format=2.5]
                S[table-format=2.5]
                S[table-format=2.5]
                S[table-format=2.4(1)]
                S[table-format=2.4(1)]
                S[table-format=2.4(1)]@{}
               }
               
\hline
\hline

   \multicolumn{1}{c}{\rule{0pt}{1.2em}$Z$~~~}      &
   \multicolumn{1}{c}{$R$~~}                        &
   \multicolumn{1}{c}{(2,2)}                        &
   \multicolumn{1}{c}{(2,3)}                        &
   \multicolumn{1}{c}{(2,4)}                        &
   \multicolumn{1}{c}{(2,5)}                        &
   \multicolumn{1}{c}{(3,3)}                        &
   \multicolumn{1}{c}{(3,4)}                        &
   \multicolumn{1}{c}{(3,5)}                        &
   \multicolumn{1}{c}{(4,4)}                        &
   \multicolumn{1}{c}{(4,5)}                        &
   \multicolumn{1}{c}{(5,5)}                        \\      
        
\hline   
                       
 110 & 6.188 &      0.3355 &      0.3551 &      0.3535 &      0.3505 &      0.4001 &      0.4113 &      0.4124 &   0.4224(1) &   0.4280(1) &   0.4315(5)  \rule{0pt}{3.6ex}   \\ 
 115 & 6.291 &      0.3456 &      0.3676 &      0.3664 &      0.3635 &      0.4162 &      0.4284 &      0.4299 &   0.4405(1) &   0.4465(1) &   0.4502(5)   \\ 
 120 & 6.391 &      0.3547 &      0.3791 &      0.3785 &      0.3756 &      0.4315 &      0.4447 &      0.4465 &   0.4578(1) &   0.4642(1) &   0.4681(4)   \\ 
 125 & 6.494 &      0.3624 &      0.3891 &      0.3893 &      0.3865 &      0.4452 &      0.4594 &      0.4616 &   0.4734(1) &   0.4801(1) &   0.4842(4)   \\ 
 130 & 6.588 &      0.3682 &      0.3970 &      0.3979 &      0.3953 &      0.4565 &      0.4715 &      0.4741 &   0.4864(1) &   0.4931(1) &   0.4974(3)   \\ 
 135 & 6.691 &      0.3717 &      0.4021 &      0.4038 &      0.4013 &      0.4645 &      0.4799 &      0.4829 &   0.4955(1) &   0.5022(1) &   0.5065(3)   \\ 
 140 & 6.785 &      0.3728 &      0.4038 &      0.4062 &      0.4040 &      0.4686 &      0.4839 &      0.4872 &      0.4999 &   0.5063(1) &   0.5106(2)   \\ 
 145 & 6.883 &      0.3716 &      0.4023 &      0.4052 &      0.4032 &      0.4689 &      0.4836 &      0.4870 &      0.4998 &   0.5054(1) &   0.5095(2)   \\ 
 150 & 6.983 &      0.3694 &      0.3986 &      0.4017 &      0.3996 &      0.4667 &      0.4803 &      0.4836 &      0.4964 &   0.5011(1) &   0.5049(2)   \\ 
 155 & 7.075 &      0.3678 &      0.3947 &      0.3975 &      0.3953 &      0.4646 &      0.4765 &      0.4794 &      0.4927 &   0.4962(1) &   0.4997(2)   \\ 
 160 & 7.170 &      0.3688 &      0.3931 &      0.3951 &      0.3924 &      0.4656 &      0.4758 &      0.4780 &      0.4921 &   0.4944(1) &   0.4975(2)   \\ 
 165 & 7.268 &      0.3737 &      0.3958 &      0.3967 &      0.3933 &      0.4723 &      0.4809 &      0.4821 &      0.4975 &   0.4987(1) &   0.5015(1)   \\ 
 170 & 7.358 &      0.3832 &      0.4040 &      0.4035 &      0.3993 &      0.4859 &      0.4931 &      0.4934 &      0.5106 &   0.5107(1) &   0.5133(1)   \\ 

\hline
\hline

\end{tabular}%

\end{table*}

%% file: table_SE_d3.tex
\begin{table*}[t]
\centering

\renewcommand{\arraystretch}{1.1}

\caption{\label{tab:se_d3} 
         Matrix elements of the self-energy operator for the $nd_{3/2}$ states evaluated with the Dirac--Coulomb wave functions.
         The notations are the same as in Table~\ref{tab:se_s}.
         }
         
\begin{tabular}{
                S[table-format=3.0]@{\quad}
                S[table-format=2.4]
                S[table-format=-2.5]
                S[table-format=2.5]
                S[table-format=2.5]
                S[table-format=2.4(1)]
                S[table-format=2.4(1)]
                S[table-format=2.4(1)]@{}
               }
               
\hline
\hline

   \multicolumn{1}{c}{\rule{0pt}{1.2em}$Z$~~~}      &
   \multicolumn{1}{c}{$R$~~}                        &
   \multicolumn{1}{c}{(3,3)}                        &
   \multicolumn{1}{c}{(3,4)}                        &
   \multicolumn{1}{c}{(3,5)}                        &
   \multicolumn{1}{c}{(4,4)}                        &
   \multicolumn{1}{c}{(4,5)}                        &
   \multicolumn{1}{c}{(5,5)}                        \\      
        
\hline   
                       
 110 & 6.188 &     -0.0064 &      0.0073 &      0.0089 &   0.0072(1) &   0.0146(1) &   0.0144(5)  \rule{0pt}{3.6ex}   \\ 
 115 & 6.291 &     -0.0009 &      0.0139 &      0.0160 &   0.0147(1) &   0.0228(1) &   0.0229(4)   \\ 
 120 & 6.391 &      0.0051 &      0.0214 &      0.0239 &   0.0231(1) &   0.0320(1) &   0.0325(4)   \\ 
 125 & 6.494 &      0.0117 &      0.0295 &      0.0326 &   0.0323(1) &   0.0421(1) &   0.0429(3)   \\ 
 130 & 6.588 &      0.0187 &      0.0380 &      0.0419 &      0.0419 &   0.0525(1) &   0.0537(3)   \\ 
 135 & 6.691 &      0.0255 &      0.0464 &      0.0512 &      0.0512 &   0.0624(1) &   0.0640(2)   \\ 
 140 & 6.785 &      0.0316 &      0.0535 &      0.0593 &      0.0590 &   0.0705(1) &   0.0723(2)   \\ 
 145 & 6.883 &      0.0362 &      0.0584 &      0.0650 &      0.0642 &   0.0755(1) &   0.0774(2)   \\ 
 150 & 6.983 &      0.0392 &      0.0609 &      0.0680 &      0.0668 &   0.0775(1) &   0.0794(2)   \\ 
 155 & 7.075 &      0.0417 &      0.0624 &      0.0695 &      0.0687 &   0.0784(1) &   0.0804(2)   \\ 
 160 & 7.170 &      0.0455 &      0.0652 &      0.0719 &      0.0721 &      0.0812 &   0.0834(1)   \\ 
 165 & 7.268 &      0.0514 &      0.0710 &      0.0772 &      0.0789 &      0.0877 &   0.0901(1)   \\ 
 170 & 7.358 &      0.0600 &      0.0805 &      0.0862 &      0.0893 &      0.0985 &   0.1009(1)   \\ 

\hline
\hline

\end{tabular}%

\end{table*}

%% file: table_SE_d5.tex
\begin{table*}[t]
\centering

\renewcommand{\arraystretch}{1.1}

\caption{\label{tab:se_d5} 
         Matrix elements of the self-energy operator for the $nd_{5/2}$ states evaluated with the Dirac--Coulomb wave functions.
         The notations are the same as in Table~\ref{tab:se_s}.
         }
         
\begin{tabular}{
                S[table-format=3.0]@{\quad}
                S[table-format=2.4]
                S[table-format=2.5]
                S[table-format=2.5]
                S[table-format=2.5]
                S[table-format=2.4(1)]
                S[table-format=2.4(1)]
                S[table-format=2.4(1)]@{}
               }
               
\hline
\hline

   \multicolumn{1}{c}{\rule{0pt}{1.2em}$Z$~~~}      &
   \multicolumn{1}{c}{$R$~~}                        &
   \multicolumn{1}{c}{(3,3)}                        &
   \multicolumn{1}{c}{(3,4)}                        &
   \multicolumn{1}{c}{(3,5)}                        &
   \multicolumn{1}{c}{(4,4)}                        &
   \multicolumn{1}{c}{(4,5)}                        &
   \multicolumn{1}{c}{(5,5)}                        \\      
        
\hline   
                       
 110 & 6.188 &      0.0699 &      0.0672 &      0.0642 &   0.0783(1) &   0.0786(1) &   0.0825(5)  \rule{0pt}{3.6ex}   \\ 
 115 & 6.291 &      0.0722 &      0.0695 &      0.0665 &   0.0812(1) &   0.0816(1) &   0.0856(4)   \\ 
 120 & 6.391 &      0.0745 &      0.0719 &      0.0688 &      0.0842 &   0.0846(1) &   0.0888(4)   \\ 
 125 & 6.494 &      0.0769 &      0.0743 &      0.0711 &      0.0871 &   0.0876(1) &   0.0921(3)   \\ 
 130 & 6.588 &      0.0793 &      0.0767 &      0.0735 &      0.0902 &   0.0908(1) &   0.0954(3)   \\ 
 135 & 6.691 &      0.0817 &      0.0792 &      0.0758 &      0.0933 &   0.0939(1) &   0.0988(2)   \\ 
 140 & 6.785 &      0.0842 &      0.0818 &      0.0783 &      0.0965 &   0.0972(1) &   0.1023(2)   \\ 
 145 & 6.883 &      0.0868 &      0.0845 &      0.0809 &      0.0999 &   0.1008(1) &   0.1061(2)   \\ 
 150 & 6.983 &      0.0896 &      0.0874 &      0.0837 &      0.1036 &   0.1046(1) &   0.1102(2)   \\ 
 155 & 7.075 &      0.0926 &      0.0906 &      0.0868 &      0.1076 &   0.1088(1) &   0.1147(1)   \\ 
 160 & 7.170 &      0.0958 &      0.0940 &      0.0902 &      0.1120 &      0.1134 &   0.1196(1)   \\ 
 165 & 7.268 &      0.0991 &      0.0977 &      0.0938 &      0.1166 &      0.1183 &   0.1249(1)   \\ 
 170 & 7.358 &      0.1026 &      0.1014 &      0.0976 &      0.1215 &      0.1234 &   0.1304(1)   \\ 

\hline
\hline

\end{tabular}%

\end{table*}

%% file: table_SE_170.tex
\begin{table}[t]
\centering

\renewcommand{\arraystretch}{1.1}

\caption{\label{tab:se_170} 
         Self-energy correction for the lowest-energy levels with the different values of $\kappa$ for $Z=170$. 
         The results are presented in terms of the function~$F_{nn}$ defined by Eq.~(\ref{eq:F}). 
         }
         
\resizebox{\columnwidth}{!}{%
\begin{tabular}{@{}
                c
                S[table-format=-2.5]
                S[table-format=-2.4(1)]
                S[table-format=-1.4]
                S[table-format=-2.4]
                S[table-format=-2.4]@{}
               }
               
\hline
\hline

   \multicolumn{1}{c}{Term\rule{0pt}{2.6ex}~~~}   &
   \multicolumn{1}{c}{~~~$1s$}                 &
   \multicolumn{1}{c}{~$2p_{1/2}$}             &
   \multicolumn{1}{c}{~~$2p_{3/2}$}            &
   \multicolumn{1}{c}{~~~$3d_{3/2}$}           &
   \multicolumn{1}{c}{~~~$3d_{5/2}$}           \\      
        
\hline   
                       
 0-pot.  &  -4.4324  &  -39.6671     &  -1.3726  &  -3.4590     &  -3.0276  \rule{0pt}{3.1ex}\\
 
 1-pot.  &   6.3439  &   48.2165     &   0.6950  &   1.8633     &   1.6695   \\
    
 2-pot.  &   1.2913  &    6.5584     &   0.2228  &   0.5355     &   0.5091   \\
 
 M-pot.  &   0.6803  &    6.4050(2)  &   0.8381  &   1.1202     &   0.9516   \\[0.5mm]    
    
\hline
    
 Total~  &   3.8831  &   21.5128(2)  &   0.3832  &   0.0600     &   0.1026   \rule{0pt}{3.1ex}\\ 

\hline
\hline

\end{tabular}%
}

\end{table}

%% file: table_WK_s.tex
\begin{table*}[t]
\centering

\renewcommand{\arraystretch}{1.1}

\caption{\label{tab:wk_s} 
         Matrix elements of the Wichmann--Kroll potential for the $ns$ states evaluated with the Dirac--Coulomb wave functions.
         The notations are the same as in Table~\ref{tab:se_s}.
         }
         
\resizebox{\textwidth}{!}{%
\begin{tabular}{
                S[table-format=3.0]@{\quad}
                S[table-format=2.4]
                S[table-format=2.5]@{\!}
                S[table-format=2.5]@{\!}
                S[table-format=2.5]@{\!}
                S[table-format=2.5]@{\!}
                S[table-format=2.5]@{\!}
                S[table-format=2.5]@{\!}
                S[table-format=2.5]@{\!}
                S[table-format=2.5]@{\!}
                S[table-format=2.5]@{\!}
                S[table-format=2.5]@{\!}
                S[table-format=2.5]@{\!}
                S[table-format=2.5]@{\!}
                S[table-format=2.5]@{\!}
                S[table-format=2.5]@{\!}
                S[table-format=2.5]@{}
               }
               
\hline
\hline

   \multicolumn{1}{c}{\rule{0pt}{1.2em}$Z$~~~}      &
   \multicolumn{1}{c}{$R$~~}                        &
   \multicolumn{1}{c}{(1,1)}                        &
   \multicolumn{1}{c}{(1,2)}                        &
   \multicolumn{1}{c}{(1,3)}                        &
   \multicolumn{1}{c}{(1,4)}                        &
   \multicolumn{1}{c}{(1,5)}                        &
   \multicolumn{1}{c}{(2,2)}                        &
   \multicolumn{1}{c}{(2,3)}                        &
   \multicolumn{1}{c}{(2,4)}                        &
   \multicolumn{1}{c}{(2,5)}                        &
   \multicolumn{1}{c}{(3,3)}                        &
   \multicolumn{1}{c}{(3,4)}                        &
   \multicolumn{1}{c}{(3,5)}                        &
   \multicolumn{1}{c}{(4,4)}                        &
   \multicolumn{1}{c}{(4,5)}                        &
   \multicolumn{1}{c}{(5,5)}                        \\      
        
\hline   
                       
 110 & 6.188 &      0.0382 &      0.0463 &      0.0455 &      0.0444 &      0.0436 &      0.0583 &      0.0575 &      0.0563 &      0.0552 &      0.0569 &      0.0556 &      0.0547 &      0.0544 &      0.0535 &      0.0525  \rule{0pt}{3.6ex}   \\ 
 115 & 6.291 &      0.0460 &      0.0569 &      0.0556 &      0.0541 &      0.0529 &      0.0734 &      0.0723 &      0.0704 &      0.0690 &      0.0712 &      0.0694 &      0.0680 &      0.0677 &      0.0663 &      0.0649   \\ 
 120 & 6.391 &      0.0558 &      0.0705 &      0.0686 &      0.0663 &      0.0647 &      0.0935 &      0.0916 &      0.0888 &      0.0867 &      0.0899 &      0.0873 &      0.0852 &      0.0847 &      0.0827 &      0.0807   \\ 
 125 & 6.494 &      0.0684 &      0.0881 &      0.0851 &      0.0818 &      0.0795 &      0.1201 &      0.1170 &      0.1128 &      0.1097 &      0.1142 &      0.1102 &      0.1072 &      0.1064 &      0.1035 &      0.1007   \\ 
 130 & 6.588 &      0.0847 &      0.1110 &      0.1062 &      0.1014 &      0.0981 &      0.1552 &      0.1501 &      0.1438 &      0.1392 &      0.1456 &      0.1396 &      0.1352 &      0.1339 &      0.1297 &      0.1256   \\ 
 135 & 6.691 &      0.1057 &      0.1403 &      0.1327 &      0.1257 &      0.1209 &      0.2008 &      0.1924 &      0.1829 &      0.1761 &      0.1849 &      0.1759 &      0.1695 &      0.1675 &      0.1614 &      0.1555   \\ 
 140 & 6.785 &      0.1330 &      0.1775 &      0.1656 &      0.1554 &      0.1485 &      0.2585 &      0.2448 &      0.2306 &      0.2208 &      0.2327 &      0.2195 &      0.2102 &      0.2071 &      0.1984 &      0.1901   \\ 
 145 & 6.883 &      0.1679 &      0.2228 &      0.2045 &      0.1899 &      0.1803 &      0.3278 &      0.3062 &      0.2856 &      0.2716 &      0.2874 &      0.2685 &      0.2555 &      0.2509 &      0.2388 &      0.2273   \\ 
 150 & 6.983 &      0.2119 &      0.2755 &      0.2487 &      0.2283 &      0.2152 &      0.4056 &      0.3735 &      0.3448 &      0.3256 &      0.3463 &      0.3202 &      0.3027 &      0.2963 &      0.2801 &      0.2648   \\ 
 155 & 7.075 &      0.2662 &      0.3341 &      0.2968 &      0.2695 &      0.2522 &      0.4876 &      0.4435 &      0.4050 &      0.3798 &      0.4068 &      0.3724 &      0.3496 &      0.3411 &      0.3203 &      0.3008   \\ 
 160 & 7.170 &      0.3313 &      0.3955 &      0.3469 &      0.3116 &      0.2896 &      0.5692 &      0.5129 &      0.4638 &      0.4319 &      0.4676 &      0.4240 &      0.3954 &      0.3849 &      0.3591 &      0.3350   \\ 
 165 & 7.268 &      0.4083 &      0.4582 &      0.3986 &      0.3546 &      0.3273 &      0.6512 &      0.5838 &      0.5230 &      0.4839 &      0.5320 &      0.4785 &      0.4434 &      0.4310 &      0.3995 &      0.3705   \\ 
 170 & 7.358 &      0.5028 &      0.5257 &      0.4556 &      0.4019 &      0.3686 &      0.7470 &      0.6684 &      0.5936 &      0.5457 &      0.6131 &      0.5475 &      0.5044 &      0.4899 &      0.4516 &      0.4163   \\ 

\hline
\hline

\end{tabular}%
}

\end{table*}

%% file: table_WK_p1.tex
\begin{table*}[t]
\centering

\renewcommand{\arraystretch}{1.1}

\caption{\label{tab:wk_p1} 
         Matrix elements of the Wichmann--Kroll potential for the $np_{1/2}$ states evaluated with the Dirac--Coulomb wave functions.
         The notations are the same as in Table~\ref{tab:se_s}.
         }
         
\begin{tabular}{
                S[table-format=3.0]@{\quad}
                S[table-format=2.4]
                S[table-format=2.4(1)]
                S[table-format=2.5]
                S[table-format=2.5]
                S[table-format=2.5]
                S[table-format=2.5]
                S[table-format=2.5]
                S[table-format=2.5]
                S[table-format=2.5]
                S[table-format=2.5]
                S[table-format=2.5]@{}
               }
               
\hline
\hline

   \multicolumn{1}{c}{\rule{0pt}{1.2em}$Z$~~~}      &
   \multicolumn{1}{c}{$R$~~}                        &
   \multicolumn{1}{c}{(2,2)}                        &
   \multicolumn{1}{c}{(2,3)}                        &
   \multicolumn{1}{c}{(2,4)}                        &
   \multicolumn{1}{c}{(2,5)}                        &
   \multicolumn{1}{c}{(3,3)}                        &
   \multicolumn{1}{c}{(3,4)}                        &
   \multicolumn{1}{c}{(3,5)}                        &
   \multicolumn{1}{c}{(4,4)}                        &
   \multicolumn{1}{c}{(4,5)}                        &
   \multicolumn{1}{c}{(5,5)}                        \\        
        
\hline   
                       
 110 & 6.188 &      0.0243 &      0.0254 &      0.0252 &      0.0249 &      0.0266 &      0.0265 &      0.0262 &      0.0263 &      0.0260 &      0.0257  \rule{0pt}{3.6ex}   \\ 
 115 & 6.291 &      0.0356 &      0.0369 &      0.0364 &      0.0358 &      0.0384 &      0.0379 &      0.0373 &      0.0375 &      0.0369 &      0.0364   \\ 
 120 & 6.391 &      0.0531 &      0.0545 &      0.0534 &      0.0523 &      0.0561 &      0.0550 &      0.0540 &      0.0540 &      0.0530 &      0.0520   \\ 
 125 & 6.494 &      0.0813 &      0.0820 &      0.0797 &      0.0777 &      0.0832 &      0.0810 &      0.0790 &      0.0788 &      0.0769 &      0.0750   \\ 
 130 & 6.588 &      0.1278 &      0.1261 &      0.1211 &      0.1172 &      0.1253 &      0.1206 &      0.1168 &      0.1161 &      0.1124 &      0.1090   \\ 
 135 & 6.691 &      0.2061 &      0.1969 &      0.1862 &      0.1787 &      0.1896 &      0.1798 &      0.1727 &      0.1705 &      0.1638 &      0.1574   \\ 
 140 & 6.785 &      0.3396 &      0.3087 &      0.2863 &      0.2718 &      0.2838 &      0.2639 &      0.2509 &      0.2457 &      0.2337 &      0.2222   \\ 
 145 & 6.883 &      0.5613 &      0.4746 &      0.4300 &      0.4033 &      0.4077 &      0.3707 &      0.3482 &      0.3375 &      0.3171 &      0.2980   \\ 
 150 & 6.983 &      0.9069 &      0.6952 &      0.6141 &      0.5686 &      0.5456 &      0.4843 &      0.4493 &      0.4306 &      0.3996 &      0.3709   \\ 
 155 & 7.075 &      1.3973 &      0.9513 &      0.8204 &      0.7504 &      0.6723 &      0.5838 &      0.5353 &      0.5079 &      0.4660 &      0.4276   \\ 
 160 & 7.170 &      2.0218 &      1.2138 &      1.0255 &      0.9278 &      0.7748 &      0.6608 &      0.5998 &      0.5648 &      0.5130 &      0.4660   \\ 
 165 & 7.268 &      2.7592 &      1.4702 &      1.2200 &      1.0931 &      0.8673 &      0.7286 &      0.6553 &      0.6137 &      0.5526 &      0.4976   \\ 
 170 & 7.358 &   3.6306(1) &      1.7407 &      1.4199 &      1.2605 &      0.9878 &      0.8177 &      0.7291 &      0.6797 &      0.6069 &      0.5421   \\ 

\hline
\hline

\end{tabular}%

\end{table*}

%% file: table_WK_p3.tex
\begin{table*}[t]
\centering

\renewcommand{\arraystretch}{1.1}

\caption{\label{tab:wk_p3} 
         Matrix elements of the Wichmann--Kroll potential for the $np_{3/2}$ states evaluated with the Dirac--Coulomb wave functions.
         The notations are the same as in Table~\ref{tab:se_s}.
         }
         
\begin{tabular}{
                S[table-format=3.0]@{\quad}
                S[table-format=2.4]
                S[table-format=2.5]
                S[table-format=2.5]
                S[table-format=2.5]
                S[table-format=2.5]
                S[table-format=2.5]
                S[table-format=2.5]
                S[table-format=2.5]
                S[table-format=2.5]
                S[table-format=2.5]
                S[table-format=2.5]@{}
               }
               
\hline
\hline

   \multicolumn{1}{c}{\rule{0pt}{1.2em}$Z$~~~}      &
   \multicolumn{1}{c}{$R$~~}                        &
   \multicolumn{1}{c}{(2,2)}                        &
   \multicolumn{1}{c}{(2,3)}                        &
   \multicolumn{1}{c}{(2,4)}                        &
   \multicolumn{1}{c}{(2,5)}                        &
   \multicolumn{1}{c}{(3,3)}                        &
   \multicolumn{1}{c}{(3,4)}                        &
   \multicolumn{1}{c}{(3,5)}                        &
   \multicolumn{1}{c}{(4,4)}                        &
   \multicolumn{1}{c}{(4,5)}                        &
   \multicolumn{1}{c}{(5,5)}                        \\     
        
\hline   
                       
 110 & 6.188 &      0.0015 &      0.0017 &      0.0017 &      0.0017 &      0.0019 &      0.0019 &      0.0020 &      0.0020 &      0.0020 &      0.0021  \rule{0pt}{3.6ex}   \\ 
 115 & 6.291 &      0.0018 &      0.0020 &      0.0021 &      0.0021 &      0.0023 &      0.0024 &      0.0024 &      0.0024 &      0.0025 &      0.0025   \\ 
 120 & 6.391 &      0.0021 &      0.0024 &      0.0025 &      0.0025 &      0.0027 &      0.0028 &      0.0029 &      0.0029 &      0.0030 &      0.0030   \\ 
 125 & 6.494 &      0.0025 &      0.0029 &      0.0030 &      0.0030 &      0.0033 &      0.0034 &      0.0035 &      0.0035 &      0.0036 &      0.0036   \\ 
 130 & 6.588 &      0.0030 &      0.0034 &      0.0035 &      0.0036 &      0.0040 &      0.0041 &      0.0042 &      0.0043 &      0.0043 &      0.0044   \\ 
 135 & 6.691 &      0.0036 &      0.0041 &      0.0043 &      0.0043 &      0.0048 &      0.0050 &      0.0050 &      0.0052 &      0.0052 &      0.0053   \\ 
 140 & 6.785 &      0.0043 &      0.0049 &      0.0051 &      0.0052 &      0.0058 &      0.0060 &      0.0061 &      0.0062 &      0.0063 &      0.0064   \\ 
 145 & 6.883 &      0.0051 &      0.0060 &      0.0062 &      0.0062 &      0.0070 &      0.0073 &      0.0074 &      0.0076 &      0.0077 &      0.0078   \\ 
 150 & 6.983 &      0.0062 &      0.0073 &      0.0075 &      0.0076 &      0.0086 &      0.0089 &      0.0090 &      0.0093 &      0.0094 &      0.0096   \\ 
 155 & 7.075 &      0.0076 &      0.0089 &      0.0093 &      0.0094 &      0.0106 &      0.0111 &      0.0112 &      0.0116 &      0.0117 &      0.0119   \\ 
 160 & 7.170 &      0.0095 &      0.0112 &      0.0116 &      0.0117 &      0.0134 &      0.0140 &      0.0142 &      0.0146 &      0.0148 &      0.0150   \\ 
 165 & 7.268 &      0.0122 &      0.0145 &      0.0150 &      0.0151 &      0.0174 &      0.0182 &      0.0184 &      0.0190 &      0.0193 &      0.0195   \\ 
 170 & 7.358 &      0.0170 &      0.0201 &      0.0208 &      0.0209 &      0.0243 &      0.0253 &      0.0256 &      0.0265 &      0.0268 &      0.0271   \\ 

\hline
\hline

\end{tabular}%

\end{table*}

%% file: table_WK_d3.tex
\begin{table*}[t]
\centering

\renewcommand{\arraystretch}{1.1}

\caption{\label{tab:wk_d3} 
         Matrix elements of the Wichmann--Kroll potential for the $nd_{3/2}$ states evaluated with the Dirac--Coulomb wave functions.
         The notations are the same as in Table~\ref{tab:se_s}.
         }
         
\begin{tabular}{
                S[table-format=3.0]@{\quad}
                S[table-format=2.4]
                S[table-format=2.6,group-separator=]
                S[table-format=2.6,group-separator=]
                S[table-format=2.6,group-separator=]
                S[table-format=2.6,group-separator=]
                S[table-format=2.6,group-separator=]
                S[table-format=2.6,group-separator=]@{}
               }
               
\hline
\hline

   \multicolumn{1}{c}{\rule{0pt}{1.2em}$Z$~~~}      &
   \multicolumn{1}{c}{$R$~~}                        &
   \multicolumn{1}{c}{(3,3)}                        &
   \multicolumn{1}{c}{(3,4)}                        &
   \multicolumn{1}{c}{(3,5)}                        &
   \multicolumn{1}{c}{(4,4)}                        &
   \multicolumn{1}{c}{(4,5)}                        &
   \multicolumn{1}{c}{(5,5)}                        \\     
        
\hline   
                       
 110 & 6.188 &     0.00013 &     0.00015 &     0.00016 &     0.00018 &     0.00019 &     0.00020  \rule{0pt}{3.6ex}   \\ 
 115 & 6.291 &     0.00017 &     0.00020 &     0.00021 &     0.00024 &     0.00025 &     0.00027   \\ 
 120 & 6.391 &     0.00022 &     0.00027 &     0.00028 &     0.00031 &     0.00033 &     0.00036   \\ 
 125 & 6.494 &     0.00029 &     0.00035 &     0.00037 &     0.00041 &     0.00044 &     0.00047   \\ 
 130 & 6.588 &     0.00038 &     0.00045 &     0.00048 &     0.00054 &     0.00057 &     0.00061   \\ 
 135 & 6.691 &     0.00050 &     0.00059 &     0.00063 &     0.00071 &     0.00075 &     0.00080   \\ 
 140 & 6.785 &     0.00066 &     0.00078 &     0.00082 &     0.00093 &     0.00098 &     0.00105   \\ 
 145 & 6.883 &     0.00086 &     0.00102 &     0.00108 &     0.00122 &     0.00129 &     0.00138   \\ 
 150 & 6.983 &     0.00114 &     0.00135 &     0.00143 &     0.00161 &     0.00171 &     0.00182   \\ 
 155 & 7.075 &     0.00152 &     0.00181 &     0.00192 &     0.00216 &     0.00230 &     0.00244   \\ 
 160 & 7.170 &     0.00208 &     0.00247 &     0.00262 &     0.00295 &     0.00314 &     0.00333   \\ 
 165 & 7.268 &     0.00296 &     0.00351 &     0.00372 &     0.00419 &     0.00445 &     0.00472   \\ 
 170 & 7.358 &     0.00467 &     0.00552 &     0.00583 &     0.00657 &     0.00695 &     0.00735   \\ 

\hline
\hline

\end{tabular}%

\end{table*}

%% file: table_WK_d5.tex
\begin{table*}[t]
\centering

\renewcommand{\arraystretch}{1.1}

\caption{\label{tab:wk_d5} 
         Matrix elements of the Wichmann--Kroll potential for the $nd_{5/2}$ states evaluated with the Dirac--Coulomb wave functions.
         The notations are the same as in Table~\ref{tab:se_s}.
         }
         
\begin{tabular}{
                S[table-format=3.0]@{\quad}
                S[table-format=2.4]
                S[table-format=2.6,group-separator=]
                S[table-format=2.6,group-separator=]
                S[table-format=2.6,group-separator=]
                S[table-format=2.6,group-separator=]
                S[table-format=2.6,group-separator=]
                S[table-format=2.6,group-separator=]@{}
               }
               
\hline
\hline

   \multicolumn{1}{c}{\rule{0pt}{1.2em}$Z$~~~}      &
   \multicolumn{1}{c}{$R$~~}                        &
   \multicolumn{1}{c}{(3,3)}                        &
   \multicolumn{1}{c}{(3,4)}                        &
   \multicolumn{1}{c}{(3,5)}                        &
   \multicolumn{1}{c}{(4,4)}                        &
   \multicolumn{1}{c}{(4,5)}                        &
   \multicolumn{1}{c}{(5,5)}                        \\     
        
\hline   
                       
 110 & 6.188 &     0.00004 &     0.00005 &     0.00005 &     0.00005 &     0.00006 &     0.00006  \rule{0pt}{3.6ex}   \\ 
 115 & 6.291 &     0.00005 &     0.00006 &     0.00006 &     0.00007 &     0.00007 &     0.00008   \\ 
 120 & 6.391 &     0.00006 &     0.00007 &     0.00008 &     0.00008 &     0.00009 &     0.00010   \\ 
 125 & 6.494 &     0.00007 &     0.00009 &     0.00009 &     0.00011 &     0.00011 &     0.00012   \\ 
 130 & 6.588 &     0.00009 &     0.00011 &     0.00012 &     0.00013 &     0.00014 &     0.00015   \\ 
 135 & 6.691 &     0.00011 &     0.00013 &     0.00014 &     0.00016 &     0.00017 &     0.00018   \\ 
 140 & 6.785 &     0.00014 &     0.00016 &     0.00018 &     0.00020 &     0.00021 &     0.00023   \\ 
 145 & 6.883 &     0.00017 &     0.00020 &     0.00022 &     0.00025 &     0.00026 &     0.00028   \\ 
 150 & 6.983 &     0.00021 &     0.00025 &     0.00027 &     0.00031 &     0.00033 &     0.00036   \\ 
 155 & 7.075 &     0.00026 &     0.00032 &     0.00034 &     0.00039 &     0.00042 &     0.00045   \\ 
 160 & 7.170 &     0.00034 &     0.00041 &     0.00044 &     0.00050 &     0.00054 &     0.00058   \\ 
 165 & 7.268 &     0.00046 &     0.00056 &     0.00060 &     0.00069 &     0.00074 &     0.00080   \\ 
 170 & 7.358 &     0.00075 &     0.00090 &     0.00097 &     0.00110 &     0.00119 &     0.00128   \\ 

\hline
\hline

\end{tabular}%

\end{table*}

%% file: table_SE_prediction.tex
\begin{table*}[t]
\centering

\renewcommand{\arraystretch}{1.1}

\caption{\label{tab:se_prediction} 
         Self-energy correction for the $4s$ and $n=5$ states in terms of the function~$F_{nn}$ defined by Eq.~(\ref{eq:F}).
         ``Exact'' labels the results of the \textit{ab initio} calculations. 
         ``Ref.~\cite{Shabaev:2013:012513}'' and ``Mod. op.'' correspond to averaging the original and current versions 
         of the model-QED operator with the Dirac--Coulomb wave function, respectively.
         }
         
\begin{tabular}{
                S[table-format=3.0]@{\quad\,}
                l@{\quad\,}
                S[table-format=2.4]
                S[table-format=2.4]
                S[table-format=2.4]
                S[table-format=2.4]
                S[table-format=2.4]
                S[table-format=2.4]@{}
               }
               
\hline
\hline

   \multicolumn{1}{c}{\rule{0pt}{1.2em}$Z$~~~}     &
   \multicolumn{1}{c}{Approach~~~~~}                 &
   \multicolumn{1}{c}{$4s$}                        &
   \multicolumn{1}{c}{$5s$}                        &
   \multicolumn{1}{c}{$5p_{1/2}$}                  &
   \multicolumn{1}{c}{$5p_{3/2}$}                  &
   \multicolumn{1}{c}{$5d_{3/2}$}                  &
   \multicolumn{1}{c}{$5d_{5/2}$}                  \\      
        
\hline   
                       
 110 & Exact                                     &       2.498 &       2.414 &       0.783 &       0.431 &       0.014 &       0.082  \rule{0pt}{3.6ex}   \\ 
     & Mod. op.                                  &       2.495 &       2.409 &       0.774 &       0.427 &       0.004 &       0.090   \\ 
     & Ref.~\cite{Shabaev:2013:012513}$^\dagger$ &       2.492 &       2.402 &       0.774 &       0.427 &       0.004 &       0.090   \\[1mm] 

 120 & Exact                                     &       2.844 &       2.712 &       1.178 &       0.468 &       0.033 &       0.089   \\ 
     & Mod. op.                                  &       2.845 &       2.714 &       1.171 &       0.463 &       0.021 &       0.097   \\ 
     & Ref.~\cite{Shabaev:2013:012513}$^\dagger$ &       2.833 &       2.691 &       1.171 &       0.463 &       0.021 &       0.097   \\[1mm] 

 130 & Exact                                     &       3.318 &       3.110 &       1.861 &       0.497 &       0.054 &       0.095   \\ 
     & Mod. op.                                  &       3.326 &       3.122 &       1.856 &       0.492 &       0.041 &       0.104   \\[1mm] 

 140 & Exact                                     &       3.847 &       3.528 &       2.924 &       0.511 &       0.072 &       0.102   \\ 
     & Mod. op.                                  &       3.862 &       3.551 &       2.925 &       0.506 &       0.058 &       0.111   \\[1mm] 

 150 & Exact                                     &       4.216 &       3.773 &       3.886 &       0.505 &       0.079 &       0.110   \\ 
     & Mod. op.                                  &       4.234 &       3.798 &       3.915 &       0.503 &       0.066 &       0.120   \\[1mm] 

 160 & Exact                                     &       4.261 &       3.725 &       3.978 &       0.498 &       0.083 &       0.120   \\ 
     & Mod. op.                                  &       4.275 &       3.740 &       3.757 &       0.499 &       0.073 &       0.130   \\[1mm] 

 170 & Exact                                     &       4.101 &       3.511 &       3.643 &       0.513 &       0.101 &       0.130   \\ 
     & Mod. op.                                  &       4.102 &       3.508 &       3.494 &       0.519 &       0.090 &       0.142   \\[1mm]

\hline
\hline

\end{tabular}%

\vspace*{0.7mm}
{\small
$^\dagger$ The original model-QED operator was constructed using the SE matrix elements from the present work.
}

\hspace*{5mm}

\end{table*}

%% file: table_WK_prediction.tex
\begin{table*}[t]
\centering

\renewcommand{\arraystretch}{1.1}

\caption{\label{tab:wk_prediction} 
         Wichmann--Kroll contribution to the vacuum-polarization correction for the $4s$ and $n=5$ states 
         in terms of the function~$F_{nn}$ defined by Eq.~(\ref{eq:F}).
         ``Exact'' labels the results of the \textit{ab initio} calculations. 
         ``Mod. op.'' corresponds to averaging the model-QED operator with the Dirac--Coulomb wave function.
         }
         
\begin{tabular}{
                S[table-format=3.0]@{\quad\,}
                l@{\quad\,}
                S[table-format=2.4]
                S[table-format=2.4]
                S[table-format=2.4]
                S[table-format=2.5,group-separator=]
                S[table-format=2.6,group-separator=]
                S[table-format=2.6,group-separator=]@{}
               }
               
\hline
\hline

   \multicolumn{1}{c}{\rule{0pt}{1.2em}$Z$~~~}     &
   \multicolumn{1}{c}{Approach~~~~~}                 &
   \multicolumn{1}{c}{$4s$}                        &
   \multicolumn{1}{c}{$5s$}                        &
   \multicolumn{1}{c}{$5p_{1/2}$}                  &
   \multicolumn{1}{c}{$5p_{3/2}$}                  &
   \multicolumn{1}{c}{$5d_{3/2}$}                  &
   \multicolumn{1}{c}{$5d_{5/2}$}                  \\      
        
\hline   
                       
 110 & Exact      &       0.054 &       0.053 &       0.026 &      0.0021 &     0.00020 &     0.00006  \rule{0pt}{3.6ex}   \\ 
     & Mod. op.   &       0.054 &       0.053 &       0.026 &      0.0021 &     0.00020 &     0.00006   \\[1mm] 

 120 & Exact      &       0.085 &       0.081 &       0.052 &      0.0030 &     0.00036 &     0.00010   \\ 
     & Mod. op.   &       0.085 &       0.081 &       0.052 &      0.0030 &     0.00035 &     0.00010   \\[1mm] 

 130 & Exact      &       0.134 &       0.126 &       0.109 &      0.0044 &     0.00061 &     0.00015   \\ 
     & Mod. op.   &       0.134 &       0.126 &       0.109 &      0.0044 &     0.00060 &     0.00015   \\[1mm] 

 140 & Exact      &       0.207 &       0.190 &       0.222 &      0.0064 &     0.00105 &     0.00023   \\ 
     & Mod. op.   &       0.207 &       0.190 &       0.222 &      0.0064 &     0.00103 &     0.00023   \\[1mm] 

 150 & Exact      &       0.296 &       0.265 &       0.371 &      0.0096 &     0.00182 &     0.00036   \\ 
     & Mod. op.   &       0.296 &       0.264 &       0.374 &      0.0096 &     0.00179 &     0.00035   \\[1mm] 

 160 & Exact      &       0.385 &       0.335 &       0.466 &      0.0150 &     0.00333 &     0.00058   \\ 
     & Mod. op.   &       0.384 &       0.333 &       0.442 &      0.0150 &     0.00327 &     0.00058   \\[1mm] 

 170 & Exact      &       0.490 &       0.416 &       0.542 &      0.0271 &     0.00735 &     0.00128   \\ 
     & Mod. op.   &       0.487 &       0.412 &       0.523 &      0.0270 &     0.00721 &     0.00127   \\[1mm]

\hline
\hline

\end{tabular}%

\hspace*{5mm}

\end{table*}

%% file: table_SE_alkali.tex
\begin{table}[t]
\centering

\renewcommand{\arraystretch}{1.1}

\caption{\label{tab:se_alkali} 
         Self-energy correction for the valence $ns$ electrons in the alkali-metal-like configurations
         ${\rm [Ne]}3s$, ${\rm [Ne]}3s^2 3p^6 4s$, and ${\rm [Ne]}3s^2 3p^6 3d^{10} 4s^2 4p^6 5s$
         in terms of the function~$F_{nn}$ defined by Eq.~(\ref{eq:F}).
         See the text for details.
         }
         
\begin{tabular}{
                S[table-format=3.0]@{\quad\,}
                l@{\quad\,}
                S[table-format=2.4]
                S[table-format=2.4]
                S[table-format=2.4]@{}
               }
               
\hline
\hline

   \multicolumn{1}{c}{\rule{0pt}{1.2em}$Z$~~~}     &
   \multicolumn{1}{c}{Approach~~~~~}                 &
   \multicolumn{1}{c}{$3s$}                        &
   \multicolumn{1}{c}{$4s$}                        &
   \multicolumn{1}{c}{$5s$}                        \\      
        
\hline   
                       
 110 & Exact                                     &       2.249 &       1.925 &       1.389  \rule{0pt}{3.6ex}   \\ 
     & Mod. op.                                  &       2.251 &       1.927 &       1.393   \\ 
     & Ref.~\cite{Shabaev:2013:012513}$^\dagger$ &       2.244 &       1.923 &       1.390   \\ 
     & H-like                                    &       2.601 &       2.498 &       2.414   \\[1mm] 

 120 & Exact                                     &       2.640 &       2.242 &       1.650   \\ 
     & Mod. op.                                  &       2.642 &       2.247 &       1.656   \\ 
     & Ref.~\cite{Shabaev:2013:012513}$^\dagger$ &       2.633 &       2.241 &       1.652   \\ 
     & H-like                                    &       3.017 &       2.844 &       2.712   \\[1mm] 

 130 & Exact                                     &       3.193 &       2.670 &       1.985   \\ 
     & Mod. op.                                  &       3.194 &       2.677 &       1.993   \\ 
     & H-like                                    &       3.611 &       3.318 &       3.110   \\[1mm] 

 140 & Exact                                     &       3.862 &       3.154 &       2.349   \\ 
     & Mod. op.                                  &       3.863 &       3.162 &       2.357   \\ 
     & H-like                                    &       4.325 &       3.847 &       3.528   \\[1mm] 

 150 & Exact                                     &       4.433 &       3.516 &       2.605   \\ 
     & Mod. op.                                  &       4.433 &       3.519 &       2.608   \\ 
     & H-like                                    &       4.916 &       4.216 &       3.773   \\[1mm] 

 160 & Exact                                     &       4.666 &       3.597 &       2.644   \\ 
     & Mod. op.                                  &       4.672 &       3.597 &       2.640   \\ 
     & H-like                                    &       5.135 &       4.261 &       3.725   \\[1mm] 

 170 & Exact                                     &       4.613 &       3.479 &       2.536   \\ 
     & Mod. op.                                  &       4.635 &       3.479 &       2.528   \\ 
     & H-like                                    &       5.066 &       4.101 &       3.511   \\[1mm]

\hline
\hline

\end{tabular}%

\vspace*{0.7mm}
{\small
$^\dagger$ The original model-QED operator was constructed using the SE matrix elements from the present work.
}

\end{table}

%% file: table_SE_1s_DHFS.tex
\begin{table*}[t]
\centering

\renewcommand{\arraystretch}{1.1}

\caption{\label{tab:se_1s_DHFS} 
         Self-energy correction for the $1s$ state in superheavy elements
         in terms of the function~$F_{nn}$ defined by Eq.~(\ref{eq:F}).
         See the text for details.
         }
         
\begin{tabular}{
                l@{\quad\,}
                S[table-format=2.3(1)]
                S[table-format=2.3(1)]
                S[table-format=2.3(1)]
                S[table-format=2.3(1)]
                S[table-format=2.3(1)]
                S[table-format=2.3(2)]@{}
               }
               
\hline
\hline

   \multicolumn{1}{c}{Approach~~~~~}                  &
   \multicolumn{1}{c}{$Z=110$}                        &
   \multicolumn{1}{c}{$Z=120$}                        &
   \multicolumn{1}{c}{$Z=130$}                        &
   \multicolumn{1}{c}{$Z=140$}                        &
   \multicolumn{1}{c}{$Z=150$}                        &
   \multicolumn{1}{c}{$Z=160$}                        \\      
        
\hline   
                       
 Ref.~\cite{Cheng:1976:1943}               &     1.53(2) &     1.67(2) &     1.92(3) &     2.30(5) &     2.78(9) &    3.34(16)  \rule{0pt}{3.6ex}   \\ 
 Mod. op.                                  &        1.55 &        1.70 &        1.95 &        2.33 &        2.83 &        3.37   \\ 
 Ref.~\cite{Shabaev:2013:012513}$^\dagger$ &        1.55 &        1.70 &             &             &             &               \\ 
 H-like                                    &        1.57 &        1.73 &        1.98 &        2.37 &        2.89 &        3.46   \\[1mm] 

\hline
\hline

\end{tabular}%

\vspace*{0.7mm}
{\small
$^\dagger$ The original model-QED operator was constructed using the SE matrix elements from the present work.
}

\end{table*}